\documentclass[preprint,nofootinbib,12pt]{revtex4-1}

\pdfoutput=1

\usepackage[utf8]{inputenc}

\usepackage{fullpage}

\usepackage{amsmath}
\usepackage{amssymb}
\usepackage{latexsym}
\usepackage{color}

\usepackage{boxedminipage}

\usepackage{ifpdf}

\usepackage{slashed}

\usepackage{pgf}
\usepackage{tikz}

\usetikzlibrary{positioning}
\usetikzlibrary{decorations.pathmorphing}
\usetikzlibrary{decorations.markings}
\usetikzlibrary{snakes}

\newcommand{\dslash}{\partial\hspace{-6pt}\slash}

\newcommand{\beq}{\begin{equation}}
\newcommand{\eeq}{\end{equation}}
\newcommand{\beqa}{\begin{eqnarray}}
\newcommand{\eeqa}{\end{eqnarray}}

\newcommand{\lsim}{\mathrel{\rlap{\lower4pt\hbox{\hskip1pt$\sim$}}
    \raise1pt\hbox{$<$}}}         
\newcommand{\gsim}{\mathrel{\rlap{\lower4pt\hbox{\hskip1pt$\sim$}}
    \raise1pt\hbox{$>$}}}         

\definecolor{myblue}{rgb}{0.2,0.2,0.7}
\definecolor{myred}{rgb}{0.7,0.2,0.2}

\begin{document}

\tikzset{ 
  scalar/.style={dashed},
  scalar-ch/.style={dashed,postaction={decorate},decoration={markings,mark=at
      position .5 with {\arrow{>}}}},
  fermion/.style={postaction={decorate}, decoration={markings,mark=at
      position .5 with {\arrow{>}}}},
  gauge/.style={decorate, decoration={snake,segment length=0.2cm}},
  gauge-na/.style={decorate, decoration={coil,amplitude=4pt, segment
      length=5pt}}
}


\vspace*{.0cm}

\title{Momentum asymmetries as CP violating observables}

\author{Joshua Berger}
\email{jb454@cornell.edu}
\author{Monika Blanke}
\email{mb744@cornell.edu}
\author{Yuval Grossman}
\email{yg73@cornell.edu}
\author{Shamayita Ray}
\email{sr643@cornell.edu}

\affiliation{\vspace*{4mm}Laboratory for Elementary Particle Physics, Cornell University\\
Ithaca, NY 14853, USA\vspace*{6mm}}

\begin{abstract}
Three body decays can exhibit CP violation that arises from interfering
diagrams with different orderings of the final state particles. We
construct several momentum asymmetry observables that are accessible
in a hadron collider environment where some of the final state
particles are not reconstructed and not all the kinematic information can
be extracted. We discuss the complications that arise from the
different possible production mechanisms of the decaying particle. 
Examples involving heavy neutralino decays in supersymmetric theories and
heavy Majorana neutrino decays in Type-I seesaw models are examined.
\end{abstract}

\maketitle

\section{Introduction}

The Large Hadron Collider (LHC) experiments are accumulating data at
an exceptional rate.  They will hopefully make groundbreaking
discoveries of new particles that will alter our picture of physics at
the TeV scale and beyond.  In order to fully appreciate the
consequences of these results, the properties of all of the new
particles must be determined.  A lot of effort has been spent devising
ways to determine the masses (see e.g.~\cite{Hinchliffe:1996iu,Lester:1999tx,Bachacou:1999zb,Hinchliffe:1999zc,Barr:2003rg,Miller:2005zp,Meade:2006dw,Gjelsten:2006tg,Lester:2007fq,Gripaios:2007is,Cho:2007dh,Burns:2008cp,Blanke:2010cm})
and spins (see e.g.~\cite{Barr:2004ze,Barr:2005dz,Meade:2006dw,Cousins:2005pq,Athanasiou:2006ef,Wang:2006hk,Wang:2008sw,Cho:2008tj,MoortgatPick:2011ix,Eboli:2011bq,Melia:2011cu}) of particles at the LHC and such
analyses will surely be the first on our road to understanding any new
physics.  Once a large amount of data is accumulated and these
properties are at least somewhat understood, the next step is to
determine more challenging properties such as couplings, flavor
structure, and CP violating phases. The determination of the latter
has recently attracted increased attention
\cite{Ellis:2008hq,Deppisch:2009nj,Han:2009ra,MoortgatPick:2009jy,Christensen:2010pf,MoortgatPick:2010wp,Tattersall:2010uu,Berger:2011wh,Kittel:2011sq,He:2011ws,Bornhauser:2011ab}.
In this paper, we study new
techniques which can make the direct observation of CP violation at
the LHC feasible.

With complete generality, CP violation arises when there are complex
phases in the Lagrangian that cannot be rotated away by field
redefinitions.  We call such phases CP-odd phases and the goal is to
measure them as accurately as possible.  Given a new CP-odd phase, we
can find a sensitive process and construct a CP-violating asymmetry 
\begin{equation}
  \label{eq:cp-asym}
  \mathcal{A}_{\rm CP} = \frac{N - \overline{N}}{N+ \overline{N}},
\end{equation}
where $N$ and $\overline{N}$ are the number of observed events from
the process and its CP conjugate respectively.  Observation of
$\mathcal{A}_{\rm CP} \neq 0$ requires interference between amplitudes
with different CP-odd and CP-even phases.  The CP-even phases, which
do not change sign under CP, can arise from the dynamics of the
process.  Note that if the momenta, and possibly the helicities, of
the final state particles can be determined, then it is possible to
avoid the condition of requiring amplitudes with different CP-even
phases by looking at triple product asymmetries (see
e.\,g.~\cite{Valencia:1988it,Kamionkowski:1989vj,Kayser:1989vw,Korner:1990yx,Bensalem:2000hq,Datta:2003mj,AguilarSaavedra:2004hu,Bartl:2004jj,Langacker:2007ur,Han:2009ra}).

\begin{figure}[!ht]
  \centering
  \begin{tikzpicture}
    \draw (0,0) node[left] {$X^0_0$} -- (2,0) -- (3.414,1.414) node[right] {$X^+_1$};
    \draw[very thick,myblue] (2,0) -- (3.414,-1.414);
    \draw (4.828,0) node[right] {$X^-_2$} -- (3.414,-1.414) --
    (4.828,-2.828) node[right] {$X^0_3$};
    \draw[snake=brace] (5.6,0.22) -- node[right] {$q_{23}^2$} (5.6,-3.10);
    \draw[fill=myred] (2,0) circle (0.3cm);
    \draw[fill=myred] (3.414,-1.414) circle (0.3cm);
 \end{tikzpicture}\hspace{0.2cm}
  \begin{tikzpicture}
    \draw (0,0) node[left] {$X^0_0$} -- (2,0) -- (3.414,1.414) node[right] {$X^-_2$};
    \draw[very thick,myblue] (2,0) -- (3.414,-1.414);
    \draw (4.828,0)  node[right] {$X^+_1$} -- (3.414,-1.414) --
    (4.828,-2.828) node[right] {$X^0_3$};    
    \draw[snake=brace] (5.6,0.22) -- node[right] {$q_{13}^2$} (5.6,-3.10);

    \draw[fill=myred] (2,0) circle (0.3cm);
    \draw[fill=myred] (3.414,-1.414) circle (0.3cm);
  \end{tikzpicture}
 \caption{Example of diagrams corresponding to amplitudes with different orderings.}
  \label{fig:different-orderings}
\end{figure}
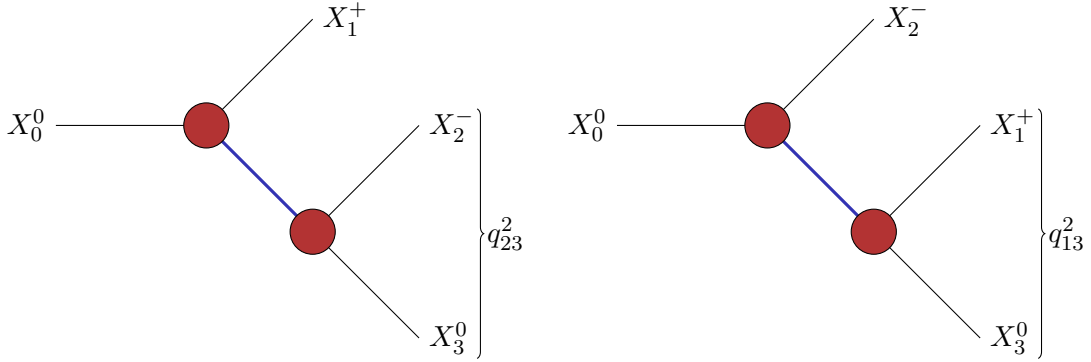
In Ref.~\cite{Berger:2011wh}, a new source of CP-even phases was
introduced.  It arises in three-body decays that can proceed with two
different final state orderings via an on-shell resonance having a finite width.
In particular, a toy model involving only scalars was considered. The
process studied was the decay $X^0_0 \to X^\pm_1 X^\mp_2 X^0_3$ via a
resonance $Y^\pm$ as illustrated in
Fig.~\ref{fig:different-orderings}. The weak phase arises from the
couplings of $Y^\pm$ to the outgoing particles. The novelty is
in the CP-even phase that arises from the different virtuality of
$Y^\pm$ in the two diagrams. Thus, once the width of $Y^\pm$ is taken
into account a differential CP asymmetry can be generated
\begin{equation}
  \label{eq:diff-cp-asym-toy-mod}
  \mathcal{A}_{\rm CP}^{\rm diff} = \frac{d\Gamma/dq_{13}^2 dq_{23}^2 -
    d\overline{\Gamma}/dq_{13}^2 dq_{23}^2}{d\Gamma/dq_{13}^2
    dq_{23}^2 + d\overline{\Gamma}/dq_{13}^2 dq_{23}^2} \; .
\end{equation}
This asymmetry is accessible via a Dalitz plot analysis.  While the CP
asymmetry constructed in this scenario can be large when considered at
specific points in phase space, the asymmetry in the total number of
events suffered a suppression when the final state particles 1 and 2
are nearly degenerate.  This suppression could be eliminated by
applying a phase space weighting, which amounted to constructing the
asymmetry
\begin{equation}
  \label{eq:phase-space-weigting}
  \mathcal{A}_{\rm CP}^{\rm PS~wgt} = \frac{\Bigl( N(q_{13}^2 > q_{23}^2) -
    N(q_{13}^2 < q_{23}^2) \Bigr) - \Bigl( \overline{N}(q_{13}^2 > q_{23}^2)  -
    \overline{N}(q_{13}^2 < q_{23}^2) \Bigr)}{N + \overline{N}} \; ,
\end{equation}
where $q_{ij}^2 = (p_i + p_j)^2$.  Still, the major practical problem
with this method is that in many cases, such as when the events
contain missing energy, some of the final state particles are not
detected and these asymmetries cannot be reconstructed.  Additional
complications arise when the particles involved are not scalars.  In
particular, since CP violating observables are sensitive only to
interference terms, it is possible to incur chiral suppression in observables.
This restriction places non-trivial limits on the set of observables
sensitive to new CP-odd phases.

In this paper, we generalize the results of \cite{Berger:2011wh} in
several ways.  We explore alternative observables based on momentum
asymmetries which are applicable in cases where $X^0_3$ escapes
detection.  We also study the conditions under which the chiral
suppression can be reduced and discuss the prospects for observing CP
asymmetries in supersymmetric models. Finally, we generalize the
observable to the case where there are two different intermediate
particles. As an example, we consider CP violation in decays of a
heavy neutrino species via $W$ and $Z$ bosons.

The main observation is as follows. When all the kinematic information
is available, momentum asymmetries can probe CP violation. The main problem
is that, in realistic scenarios, some kinematic information is not
available generically, and in particular, there is in general a large
boost so that the rest frame of the decaying particle cannot be 
reconstructed. We overcome this problem by showing that $p_T$
asymmetries are also sensitive to CP violation. More generally, as long as
we can find a direction where the boost is small (or known) we can
construct an asymmetry that can be used to probe CP violation.

The rest of this paper is structured as follows.  In
section~\ref{sec:gener-prop-moment} we introduce the momentum
asymmetry and discuss its general properties.
Section~\ref{sec:survey-observables} is devoted to a detailed survey
of observables related to this momentum asymmetry. Three different
production mechanisms for the decaying parent particle and their
consequences are discussed in detail.  In section~\ref{sec:spin} we
analyze the impact of non-zero particle spins on the asymmetries in
question. We study a supersymmetric example for concreteness.  In
section~\ref{sec:non-ident-interm} we generalize our 
study to the case where the interfering diagrams are mediated by
different intermediate resonances. A concrete example is given by
weak-scale Majorana neutrino decay.  In section~\ref{sec:conclusions}
we summarize our results and give a brief outlook.

\section{Momentum asymmetries: General properties}
\label{sec:gener-prop-moment}

To get a handle on the possibilities and set a common notation, we
introduce a minimal toy model that highlights our observables.  The
toy model is similar to that in Ref.~\cite{Berger:2011wh}.  The model
contains scalar particles $X_0^0$, $X_1^\pm$, $X_3^0$ and $Y^\pm$
charged under a $U(1)$ symmetry, with no additional symmetries.  We
will assume that $X_1$ and $X_3$ are massless, with the remaining
masses:
\begin{equation}
  \label{eq:toy-masses}
  m_0 \equiv m_{X_0} \;, \qquad m \equiv m_Y \; , \qquad m_0 > m \; .
\end{equation}
The interactions are given by
\begin{equation}
  \label{eq:lint}
  - \mathcal{L}_{\rm int} = a X_0^0 X_1^+ Y^- + b  X_3^0 X_1^+ Y^- +
  {\rm h.c.} \; ,
\end{equation}
allowing for the decays $X_0^0 \to X_1^\pm Y^\mp \to X_1^+ X_1^- X_3^0$. In
general there is a relative phase between $a$ and $b$, which we define as
$\phi_{ab}=\arg(ab^*)$. Any CP asymmetry that we construct depends on
it, and scales as $\sin2\phi_{ab}$. 

\begin{figure}[tb]
  \centering
  \begin{tikzpicture}
    \draw[scalar] (0,0) node[left] {$X_0^0$} -- (2,0);
    \draw[scalar] (2,0) -- (3.414,1.414) node [above right] {$X_1^+$};
    \draw[scalar] (2,0) -- node[above right] {$Y^-$} (3.414,-1.414);
    \draw[scalar] (3.414,-1.414) -- (4.828,0) node[right] {$X_1^-$} ;
    \draw[scalar] (3.414,-1.414) -- (4.828,-2.828) node[below right] {$X_3^0$} ;
 \end{tikzpicture}\hspace{2cm}
   \begin{tikzpicture}
    \draw[scalar] (0,0) node[left] {$X_0^0$} -- (2,0);
    \draw[scalar] (2,0) -- (3.414,1.414) node [above right] {$X_1^-$};
    \draw[scalar] (2,0) -- node[above right] {$Y^+$} (3.414,-1.414);
    \draw[scalar] (3.414,-1.414) -- (4.828,0) node[right] {$X_1^+$} ;
    \draw[scalar] (3.414,-1.414) -- (4.828,-2.828) node[below right] {$X_3^0$};
 \end{tikzpicture}
  \caption{Diagrams for the decay of $X_0$ in the toy model.}
  \label{fig:diag-decay}
\end{figure}
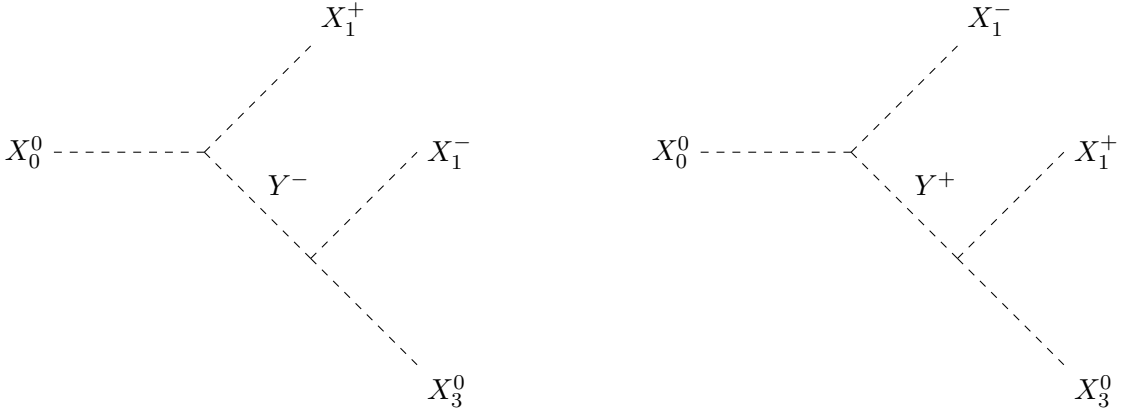
Eventually, we will need to assume some production mechanism for
$X^0_0$, but we defer a complete discussion of production to later
sections. The decay that we consider is illustrated in
Fig.~\ref{fig:diag-decay}.  Under the assumption of massless final
state particles, CP asymmetries in this decay depend on only two
parameters in addition to the phase $\phi_{ab}$:
\begin{equation}
  \label{eq:toy-params}
  \hat{m} \equiv \frac{m}{m_0} \; , \qquad \hat{\Gamma} \equiv \frac{\Gamma}{m_0} \; .
\end{equation}
Here $\Gamma \equiv \Gamma_Y$ is the width of the intermediate
particle $Y^\pm$.  The overall scale $m_0$ is only relevant when we
consider production in addition to decay.

This toy model has the feature that, after integrating over phase
space, the decay $X_0^0 \to X_1^+ X_1^- X_3^0$ is self-conjugate under
CP.  Thus, the counting experiment asymmetry of eq.~\eqref{eq:cp-asym}
vanishes trivially if $N$ is just the total number of events.  On the
other hand, we can construct an asymmetry like that in
eq.~\eqref{eq:phase-space-weigting}.  This asymmetry is generally
non-zero and we will call this the ideal asymmetry, since all the
effects that we discuss below suppress it. If the kinematics of the
decay can be reconstructed, then the strategy for looking for CP
violation is now clear: use the full knowledge of the decay to
construct the ideal asymmetry.

\begin{figure}
\centering
\includegraphics[height=0.31\textwidth]{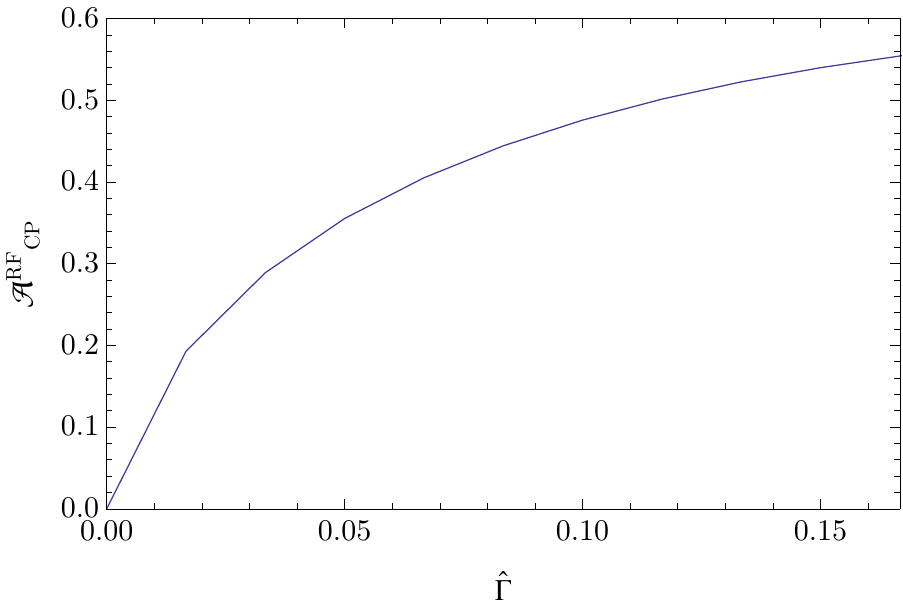}\;
\includegraphics[height=0.31\textwidth]{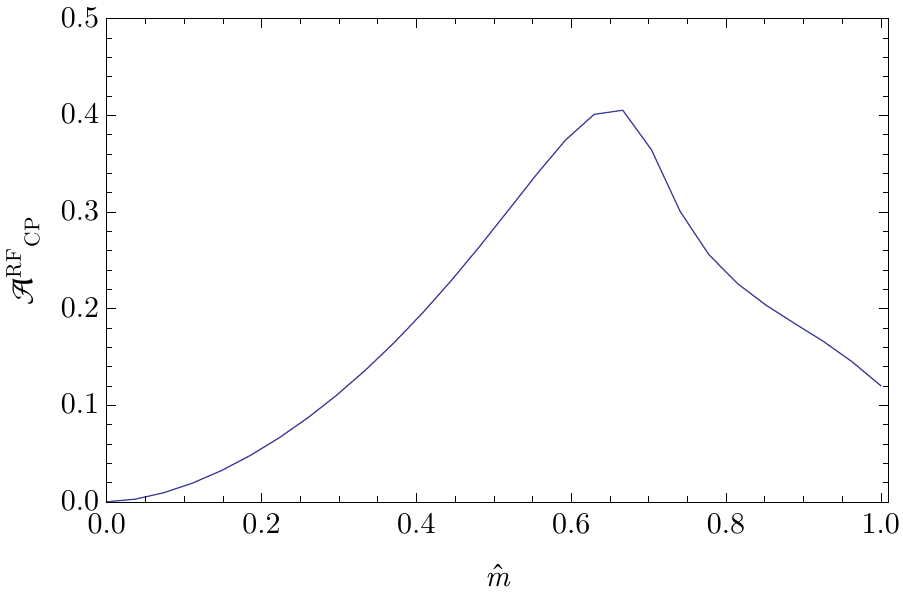}
\caption{The ideal asymmetry \eqref{eq:phase-space-weigting} as a function of $\hat\Gamma$ for fixed $\hat m =2/3$, and as a function of $\hat m$ for fixed $\hat\Gamma=1/10\,\hat{m}$.\label{fig:hatGammam}}
\end{figure}

We would like to choose some benchmark parameters to study since the
analytic expression for many of the observable we will discuss below
are complicated and unilluminating.  {The width $\hat\Gamma$ cannot be too
large or else the Breit-Wigner approximation breaks down. (While the
breakdown of the Breit-Wigner approximation affects the formulae and
quantitative results for the CP asymmetry, the qualitative result is
unaffected.) On the other hand, it cannot be too small or else the
asymmetry becomes suppressed, as it is proportional to $\hat\Gamma$ in the limit $\hat\Gamma\to 0$.  We thus choose a large, but not too large value of $\hat\Gamma$. For
$\hat{m}$ we found that the effects are largest around $\hat{m}=2/3$, when the relevant interference region lies inside the physical phase space (Dalitz plot). The dependence of the ideal asymmetry \eqref{eq:phase-space-weigting} on the two parameters $\hat{\Gamma}$ and $\hat{m}$ is shown in Fig.\ \ref{fig:hatGammam}.} We choose a maximal CP violating phase.  To summarize, the
parameters we choose are
\begin{equation}
  \label{eq:params-toy}
  \hat{m} = \frac{2}{3} \; ,\qquad \hat{\Gamma} = \frac{2}{30} \;
,\qquad \phi_{ab}=\frac{\pi}{4} \; .
\end{equation}

It is instructive to present the asymmetries in term of
momentum asymmetries. Working in the $X_0$ rest frame, we have 
\beq
q_{13}^2 = (p_0 - p_-)^2 = m_0^2 - 2 m_0 p_-^{\rm RF} \; ,
\qquad q_{23}^2 = m_0^2 - 2 m_0 p_+^{\rm RF} \; .
\eeq  
Thus, for the case we are considering,
eq.~\eqref{eq:phase-space-weigting} can be reduced to
\begin{equation}
  \label{eq:asym-ideal} \mathcal{A}_{\rm CP}^{\rm RF} = \frac{N(p_-^{\rm
  RF} > p_+^{\rm RF}) - N(p_+^{\rm RF} > p_-^{\rm RF})}{N} \; .
\end{equation}
The ideal asymmetry is in this way written as a momentum asymmetry.
For our choice of parameters, eq.~\eqref{eq:params-toy},  we find
\begin{equation}
  \label{eq:asym-rf-max}
  \mathcal{A}_{\rm CP}^{\rm RF} = 0.405 \; .
\end{equation}

The practical problem is that the ideal observable cannot be measured
in hadron collider experiments.  There are three sources of suppression
that can enter the determination of a realistic observable: loss of
kinematic information, combinatorics, and energy
smearing effects. We discuss each of them in turn.

The most difficult source of suppression to deal with is that of
kinematic information loss.  If all three particles in the decay can
be measured, then there is no issue and the ideal asymmetry
\eqref{eq:asym-ideal} can be measured up to energy smearing. In what
follows we make the assumption that the neutral particle $X_3^0$ is
stable and escapes the detector.  In this case, it is not possible to
reconstruct the rest frame momenta of the charged particles on an
event-by-event basis.  New observables will need to be devised to make
use of the available data.  

Energy smearing due to detector resolution can be a large effect if
the detector resolution is not sufficient to probe the width of the
resonance.  In Ref.~\cite{Berger:2011wh}, the largest asymmetry was
seen to come from a region of phase space of order $\Gamma$ away from
the point where both resonances are on shell.  If the resolution is
insufficient to distinguish momenta in this region, then there will be
a suppression of the net asymmetry.

Combinatorics causes a suppression if we cannot correctly determine
which particles came from the same mother particle.  In that case, the
best that we can do is to calculate observables using all possible
combinations of momenta.  When the momenta forming the observable have
small correlation, as will be the case for certain variables, there
will only be a small suppression.  On the other hand, this can be a
non-trivial effect in some cases and needs to be taken into account.

We now examine ways to approximate \eqref{eq:asym-ideal} in three
scenarios within the confines of a realistic detector scenario.

\section{Survey of observables}
\label{sec:survey-observables}

To make progress at this point we must assume a production mechanism
for the mother particles in a collider environment.  There are three
possibilities which we will discuss in turn: resonant production, pair
production, and production via decay.  A fourth possibility,
associated production, generally works similarly to pair production
and we do not discuss it separately.

The goal in studying each of these scenarios is to best reproduce the
ideal asymmetry \eqref{eq:asym-ideal}.  It will not be possible to
obtain quite so large an asymmetry in realistic scenarios, but we will
demonstrate that the suppressions due to energy smearing,
combinatorics, and lost kinematic information can be overcome and
viable observables can still be constructed.

Since the lower energy runs of the LHC are unlikely to generate enough
statistics to perform precision studies, we consider proton-proton
collisions at 14 TeV throughout.  All collider results are obtained
using events generated by MadGraph5~\cite{MG5}.  The model input for
MadGraph was generated using FeynRules
1.6.0~\cite{Christensen:2008py}.  The events are studied at parton
level with no cuts unless otherwise specified. We 
generated only signal events and our sample size is $10^5$ events,
which leads to a statistical uncertainty of order few per thousand.  Once we
consider collider observables, the mass of the $X_0$ particle becomes
relevant.  Throughout this section, we assume $m_0 = 400~{\rm GeV}$
with all other relative masses as in Section
\ref{sec:gener-prop-moment}. 

{A more realistic analysis including initial and final state radiation, showering and hadronization effects and full detector simulation is beyond the scope of our analysis, given that we focus mainly on an unrealistic toy model. These effects generally give rise to a further smearing of particle momenta and cause additional combinatoric uncertainties. While a fully realistic simulation should take these effects into account, we do not expect them to have a significant impact on our results as we found the effects of energy smearing and combinatorics to be rather small.}

\subsection{Resonant production}
\label{sec:resonant-production}

The simplest possibility is that the mother particle is produced as a
resonance. To study this case, we add a vertex
\begin{equation}
  \label{eq:resonance-vertex}
  \mathcal{L} \supset - \lambda X^0_0 \overline{q} q^\prime  \; ,
\end{equation}
and we analyze the process
\begin{equation}
  \label{eq:resonance-process}
  p p \to X^0_0 \to X^+_1 X^-_1 X^0_3 \; .
\end{equation}

In this case, the $X^0_3$ is not detected and its momentum cannot be
determined without further information about the spectrum of the
model.  We can, however, approximate the asymmetry
\eqref{eq:asym-ideal} using information from only the $X_1$ particles.
In the parton center of mass (CM) frame, $X^0_0$ will be produced at rest.  The
boost from the parton rest frame to the proton rest frame is to very
good approximation longitudinal, so any transverse variables can be
compared on an event by event basis.  Thus, we are led to examine a
$p_T$ asymmetry \cite{Han:2009ra,Schmidt:1992et}
\begin{equation}
  \label{eq:asym-resonance}
  \mathcal{A}_{\rm CP}^{p_T} = \frac{N(p_{T,-} > p_{T,+}) - N(p_{T,+} > p_{T,-})}{N} \; .
\end{equation}
Using our representative numbers, eq.~\eqref{eq:params-toy}, we find
that the asymmetry of eq.~\eqref{eq:asym-resonance} is given by
\begin{equation}
  \label{eq:asym-resonance-max}
  \mathcal{A}_{\rm CP}^{p_T} = 0.209 \; ,
\end{equation}
indicating a suppression of about a factor of $1/2$ compared to
eq.~\eqref{eq:asym-rf-max}.  
The result is consistent with a numerical calculation in the rest frame of the $X_0$.

\begin{figure}[!tb]
  \centering
\includegraphics[height=0.35\textwidth]{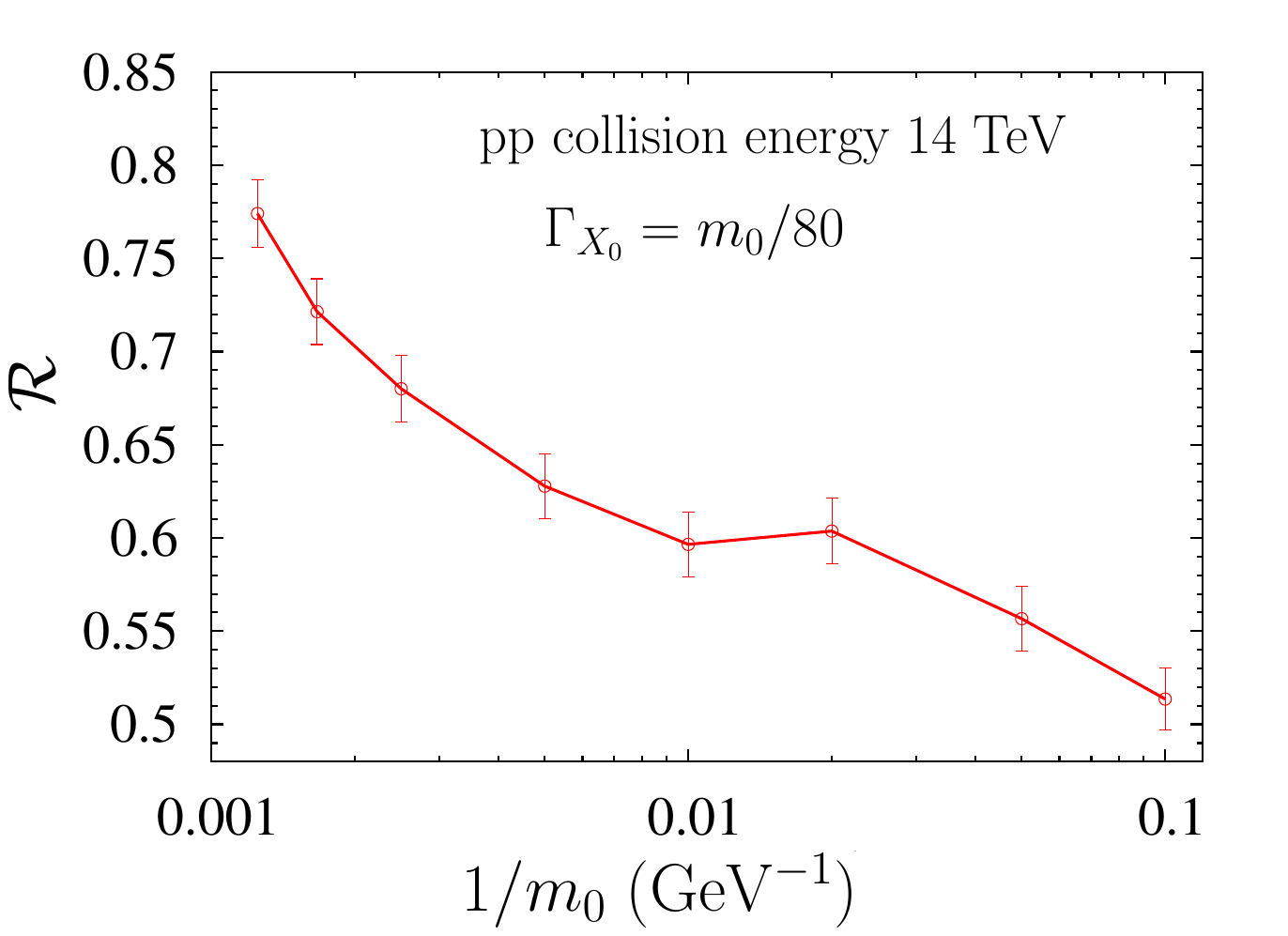}
\includegraphics[height=0.35\textwidth]{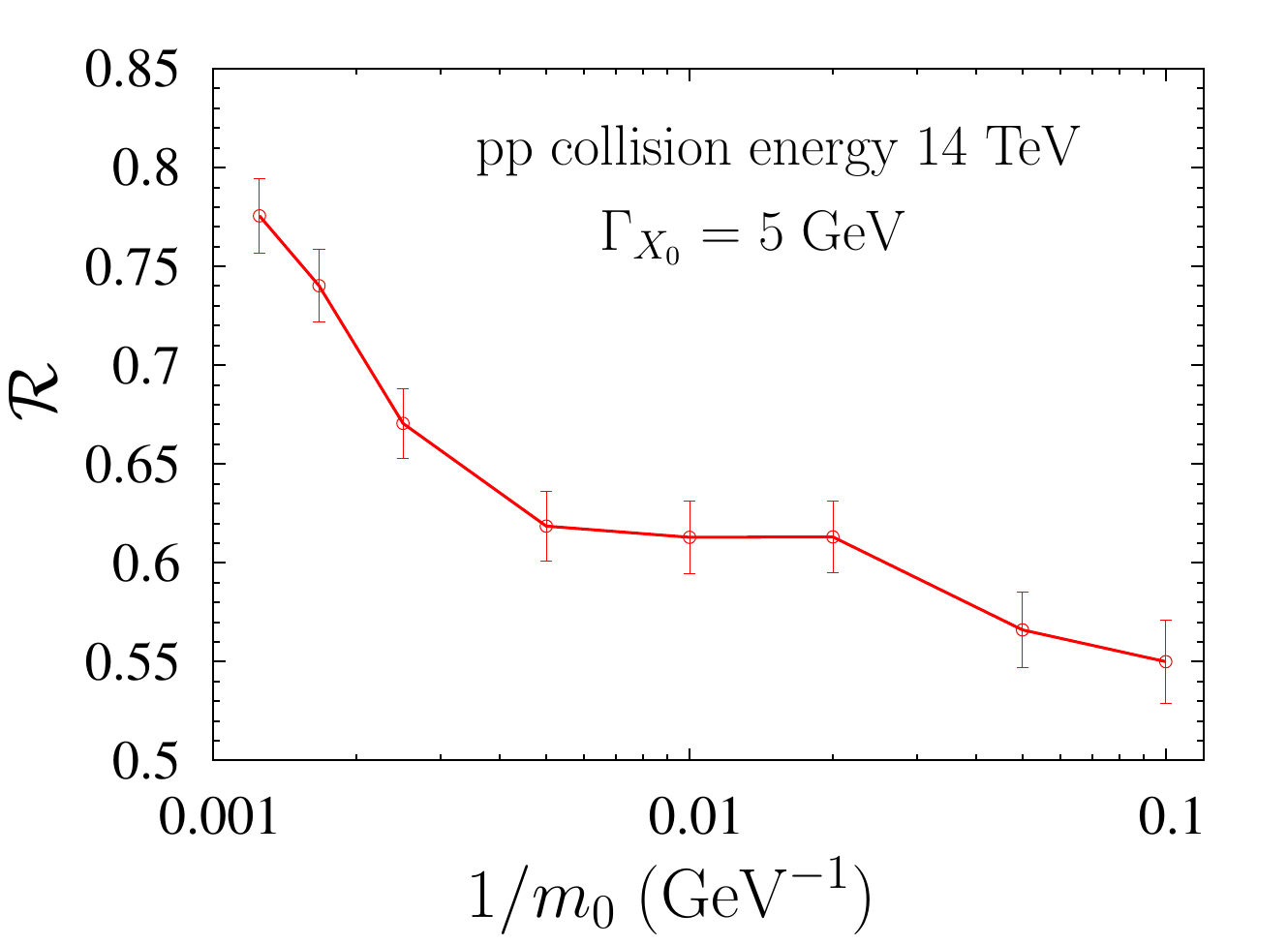}
  \caption{The ratio $\mathcal R$ of the asymmetry in the lab frame, 
eq.~\eqref{eq:asym-mom}, to that in the restframe, 
eq.~\eqref{eq:asym-resonance-max}, as a function of $m_0$.  $\hat m$,
$\hat \Gamma$ and $\phi_{ab}$ are kept the same as in
eq.~\eqref{eq:params-toy}.   
Left: We keep the width to mass ratio fixed. Right: We keep the
  width fixed.}  \label{fig:boost-dep}
\end{figure}

If we attempted to consider an asymmetry using the full
momentum of the charged particles, then the information in the $z$
direction would be washed out by the longitudinal boost.  The
momentum asymmetry
\begin{equation}
  \label{eq:asym-mom}
  \mathcal{A}_{\rm CP}^p = \frac{N(p_- > p_+) - N(p_+ > p_-)}{N} \; ,
\end{equation}
should be comparable but smaller than eq.~\eqref{eq:asym-resonance}.
This is indeed what we find:  
\begin{equation}
  \label{eq:asym-mom-num}
  \mathcal{A}_{\rm CP}^p = 0.140 \; .
\end{equation}
The amount of suppression is related to the longitudinal boost. We check this dependence by varying the mass of $X_0$,
and plot the amount of suppression due to the boost in Fig.~\ref{fig:boost-dep}.

There is only one possible choice for combinatorics in this situation,
so we next discuss detector resolution effects.  Ultimately the
charged particles $X_1$ will have to be well measured if they are to
be charge tagged.  In the SM, the stable particles that can be charge
tagged are the leptons, so for concreteness we take the energy
smearing to be that of muons at CMS~\cite{cms-muons}.  We can
parametrize the resolution by the approximation:
\begin{equation}
  \label{eq:muon-res}
  \frac{\Delta p_T}{p_T} = 0.08 \frac{p_T}{1~{\rm TeV}} \oplus 0.01 \; .
\end{equation}
On the other hand, the width of $Y$ in this scenario is
$\Gamma_Y \approx 27~{\rm GeV}$, using \eqref{eq:params-toy} and 
$m_0 = 400~{\rm GeV}$.  The smearing is much less than the width for $p_T
\lesssim 500~{\rm GeV}$, which covers almost all of the generated
events.  Smearing is thus expected to be a negligible effect in
this scenario and we have verified that this is indeed true using our
generated events.

Using transverse momentum asymmetries, we have constructed an
observable that is suppressed by only a factor of about a half compared
to the ideal asymmetry in the case of resonant production.  The
observables that we will study in the remaining cases will work on a
similar principle: we find variables that are almost invariant under
boosts in a relevant direction.

\subsection{Pair production}
\label{sec:pair-production}

The case of pair production is similar in many respects to the case of
resonant production, yet there are a few key differences.  The
asymmetry we would like to consider is still
eq.~\eqref{eq:asym-resonance}. The reason is that the pair of $X_0$s
will typically be produced near threshold and thus the boost in the
transverse direction is not large enough to wash the effect away. Yet,
the production is not exactly on threshold and there is some
significant boost in the transverse direction. This implies the
breakdown of the assumption of zero transverse momentum that we used
in the resonance production case.  In addition, there are also
combinatoric effects, but, as we will argue, they are small.

We consider production via a neutral scalar $S$, that is, $p p \to S
\to X^0_0 X^0_0$.  We assume that $S$ has a mass of $120~{\rm GeV}$
and that it couples both to quarks and to $X^0_0$s in a way that does
not violate parity.  We also let $X^0_0$ be much heavier than $S$ at
$400~{\rm GeV}$.  This assumption ensures that the $X^0_0$ will be
produced close to threshold.  The particular choice of masses should
not have a large effect on the general conclusions reached in this
section.

The most significant new effect is due to the transverse momentum of
the $X^0_0$.  This will tend to wash out the asymmetry by a factor
proportional to the typical transverse energy of $X^0_0$.  For the
choice of parameters we have made, the average $p_T$ of $X^0_0$ is
about $200~{\rm GeV}$, which is comparable to the mass scales of the
system.

We present the results of the calculation of the asymmetry in three
ways.  In the first instance, we use Monte Carlo information to
correctly associate the charged particles with each other and do not
apply any smearing, and we obtain 
\beq \label{eq:asy-pair-net}
\mathcal{A}_{\rm CP}^{p_T}=0.127 \; .
\eeq
The suppression in this case is entirely due to the fact that the
$X_0$ is not exactly at rest in the transverse direction.

\begin{figure}[!tb]
  \centering
  \begin{tikzpicture}
    \node[anchor = west] at (-8,0)
    {\includegraphics[width=0.45\textwidth]{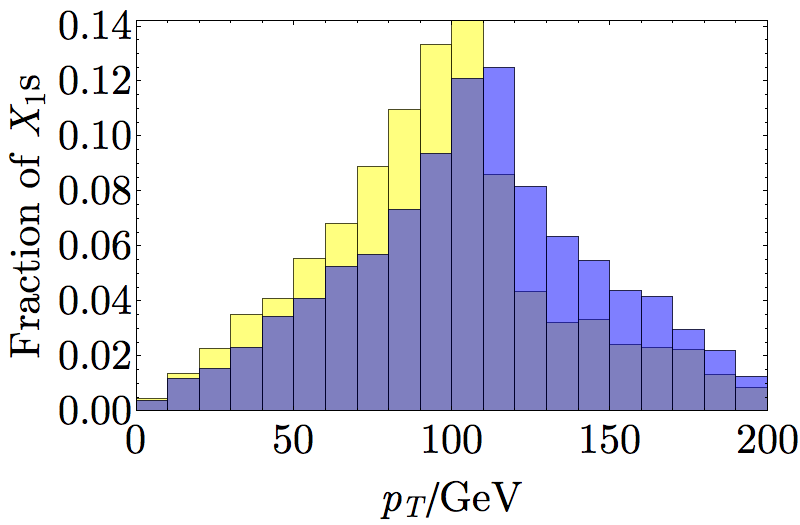}};
    \node[anchor = east] at (8,0)
    {\includegraphics[width=0.45\textwidth]{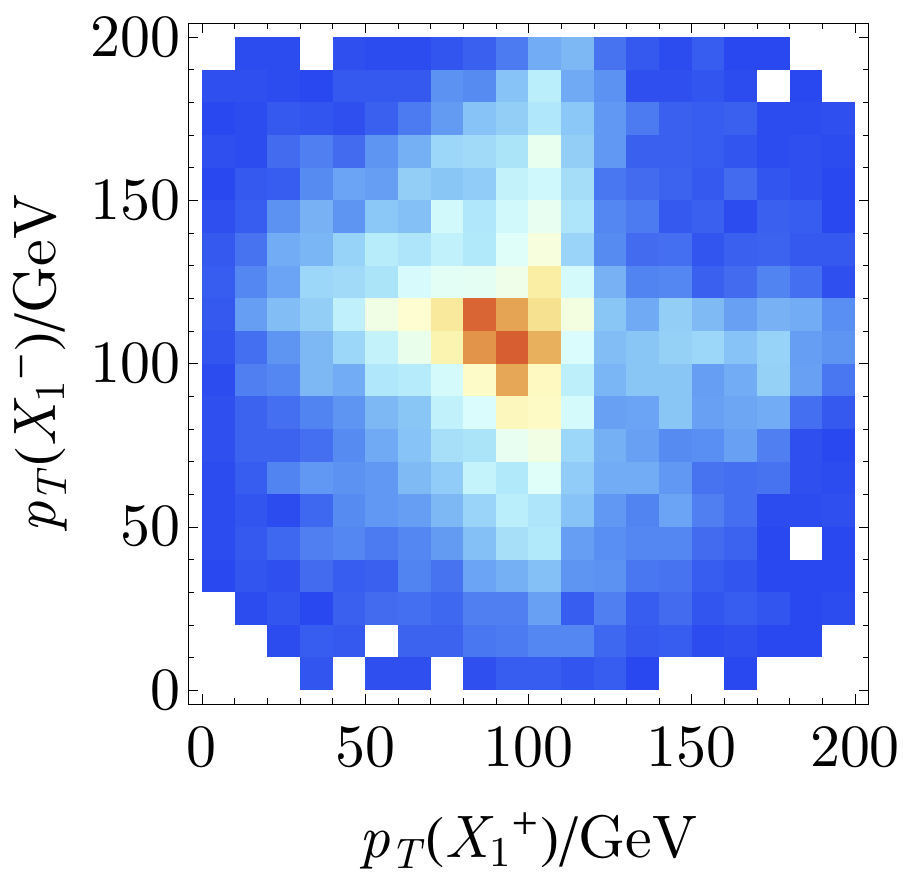}};
  \end{tikzpicture}
  \caption{Left: The $p_T$ distribution of $X_1^+$ (yellow) and $X_1^-$
    (blue) with the overlap in purple.  Right: The $p_T$ distribution of $X_1^+$ and $X_1^-$.
    Blue corresponds to a small fraction of events, while red
    corresponds to a large fraction of events (up to a maximum of
    $1.7~\%$ of events).}
  \label{fig:combinatorics-pair}
\end{figure}
The effect of combinatorics should be taken into account next.  It
will, however, be small due to the fact that the majority of the
asymmetry is generated from the difference in the average $p_T$
of the $X_1^+$ and $X_1^-$.  The correlation on an event-by-event
basis is small.  The points that exhibit such a correlation are close
to the edges of phase space where the differential rate is small.
This fact is illustrated in Fig.~\ref{fig:combinatorics-pair} using
the illustrative case of resonant production.  Even wrong pairings
will on average exhibit an asymmetry nearly as large as the correct
pairings.  After including combinatorics, we obtain a result that is
consistent with eq.~\eqref{eq:asy-pair-net} within uncertainties, confirming
that correlations between the $p_T$ of positively and negatively
charged particles are negligibly small in this case.

Smearing is expected to be a negligible effect in this case as well
and we have verified that this is the case.  Thus, the overall
suppression in the case of pair production is not too large.  The
total suppression compared to the ideal asymmetry is a factor of about
$1/4$.


\subsection{Production via decay}
\label{sec:production-via-decay}

For production via decay, the situation can be more complicated.  We
consider one of the simpler possibilities.  On top of the particles in
the pair production model, we introduce two more scalar particles
$\Phi$ and $\phi$.  We introduce couplings such that $\Phi$ are pair
produced in the process $p p \to S \to \Phi\Phi$.  Then, $\Phi$ decays
to $\phi X_0$ and $X_0$ decays as in the previous cases.  We assume
that $\Phi$ has a mass of $1~{\rm TeV}$ and that $\phi$ is massless.

We open up a new possibility here, since $\Phi$ and $\phi$ may be
colored. This provides a way to get an enhanced production
cross-section.  On the other hand, the momentum resolution for the
resulting jets is worse, which will cause a manageable suppression of
the asymmetry, as we discuss shortly.

In general, the momentum of $X_0$ will be correlated 
with the momentum of $\Phi$.  If $\Phi$ is sufficiently boosted, then
the momenta should align to good approximation, just as the momentum
of the $X_0$ aligned with the beam axis to good approximation in the
previous two cases.  We can then consider the momentum of $X_0$
particles transverse to the axis defined by the momentum of
$\Phi$.  Since we cannot determine this axis exactly, we can
approximate it by taking it to be the axis defined by the momentum of
the light particle $\phi$ that also comes out of this decay.  We thus
define the momentum
\begin{equation}
  \label{eq:jet-trans}
  p_{T,ij} \equiv \frac{|p_i \times p_j|}{|p_j|}
\end{equation}
and the asymmetry in this variable
\begin{equation}
  \label{eq:jet-trans-asym}
  \mathcal{A}_{\rm CP}^{\phi_T} = \frac{N(p_{T,-\phi} > p_{T,+\phi}) - N(p_{T,+\phi} > p_{T,-\phi})}{N} \; .
\end{equation}

The suppressions relative to the ideal case occur in a similar fashion
to the previous two cases, with the main suppression being due to the
fact that this transverse momentum is not quite invariant under the
relevant boosts of $X_0$.  In this case, however, combinatorics may be a
larger source of suppression since we now must match the charged
particles with the correct $\phi$.  Unlike in the case of pair
production, there is significant correlation between the momenta of
the $\phi$s and the corresponding $X_1$ momenta.  Furthermore, if
$\phi$ is taken to be colored, smearing will also cause a non-trivial
suppression.  To take this into account, we use the jet energy
resolution at CMS \cite{cms-jets} to smear the momenta of the $\phi$.
We parametrize the resolution as a function linear in the logarithm of
the jet $p_T$ such that for $p_T = 10~{\rm GeV}$, the resolution is
$16\,\%$ and for $p_T = 2~{\rm TeV}$, the resolution is $4\,\%$.
The results for our choice of parameter are
\begin{equation}
\mathcal{A}_{\rm CP}^{\phi_T}=0.122
\end{equation} 
when combinatorics and smearing are included.

This concludes our tour of the possible production mechanisms for the
mother particle $X_0$.  Without using relatively complex kinematic
constructions, we cannot fully reconstruct events and examine the
ideal observable.  We can, however, construct alternative observables
that have relatively small suppressions.  These observables rely on
the approximation that the $X_0$ is produced with small momentum
transverse to some axis.  

\subsection{Comparison with triple product asymmetries}

Here we compare our new observables to another class of observables,
triple products.  Such observables arise in a Lorentz invariant way
from contraction of the Levi-Civita symbol with four independent
four-vectors, that is $\epsilon^{\mu\nu\rho\sigma} p_{1\mu} p_{2\nu}
p_{3\rho} p_{4\sigma}$.  A general triple product observable is then
defined by
\begin{equation}
  \label{eq:cpv-tp}
  \mathcal{A}_{\rm CP}^{\rm T.P.} = \frac{N(\epsilon^{\mu\nu\rho\sigma} p_{1\mu} p_{2\nu} p_{3\rho}
p_{4\sigma} > 0) - N(\epsilon^{\mu\nu\rho\sigma} p_{1\mu} p_{2\nu} p_{3\rho}
p_{4\sigma} < 0)}{N} \; .
\end{equation}
If all of the $p$ vectors are momenta, then this observable is P-odd, 
so that for it to be CP-violating, it must be C-even.  In other
words, all of the final state particles must be  C
invariant or C must interchange exactly two pairs of particles.  If
these conditions are not satisfied, more complicated observables
involving the triple product can be constructed.  It is important
to note that epsilon contractions of this 
form always come with a factor of $i$ in field theory and that this
factor does not change sign under CP.  This $i$ acts as a maximal
$\pi/2$ CP-even phase, generally giving triple products an advantage
over other CP-violating observables in weakly coupled theories: they
do not have to have an additional source of CP-even phase.  If a
triple product can be constructed, then it will typically be favorable
to do so.  On the other hand, in order for such an observable to be
relevant, there need to be four independent four-vectors from which we
can construct a CP-odd contraction.  In the case of resonant
production, we are considering a $2 \to 3$ scattering process, where
incoming partons collide to produce the three particles $X_1^+$,
$X_1^-$ and $X_3^0$.  Thus, we only have three independent final state
momenta with which to construct our observable in principle and cannot
form a non-vanishing triple product \cite{Golowich:1988ig}.  We expect
momentum asymmetries to be the only ones sensitive in 
cases where the CP-violation occurs in a 3-body decay where the final
state is stable and helicities cannot be measured.

If the final state particles can
decay, then the momenta of the decay products provide information
about the helicity of their mothers and give a new set of correlated
four-vectors.  Similarly, if the helicity can be measured, there are
enough four-vectors to construct a triple product.  If the production
violates P or CP, then it is possible to construct a CP-violating
observable using triple products.  In the case of parity violation,
the production is now correlated with the helicity of the mother of
the three-body decay, the analogue of $X_0^0$, and allows for a
non-trivial triple product.  In the case of CP violation, it becomes
unclear what phase is being measured, as it will really be a
combination of the production phase and the decay phase.

\section{Spin and momentum asymmetries}
\label{sec:spin}

In general, spin is not expected to play a significant role in
determining the size of the asymmetry.  On the other hand, states with
different helicity are distinguishable quantum states and do not interfere
with one another, leading to effects on the interference term crucial
to CP violating observables.  If the couplings of the theory exhibit a
chiral structure, then observables can face a significant suppression.

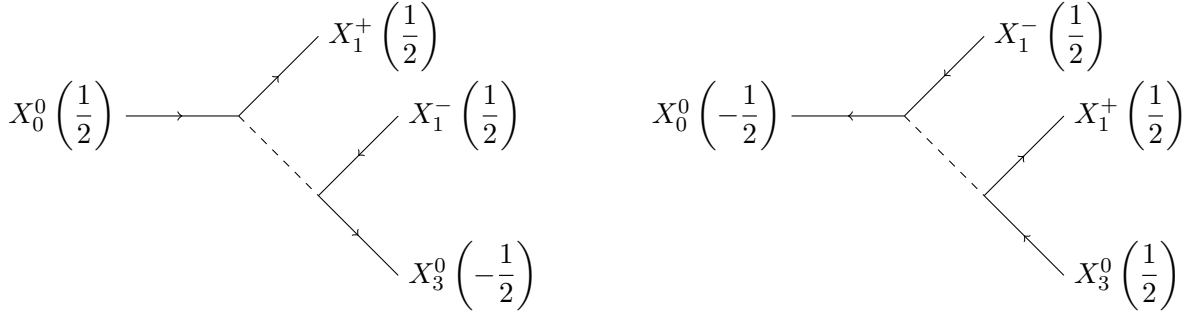
\begin{figure}[tb]
  \centering
  \begin{tikzpicture}[scale=0.75]
    \draw[fermion] (0,0) node[left] {\small $X^0_0\left(\dfrac{1}{2}\right)$} -- (2,0);
    \draw[fermion] (2,0) -- (3.414,1.414) node[right] {\small $X^+_1\left(\dfrac{1}{2}\right)$};
    \draw[scalar] (2,0) -- (3.414,-1.414);
    \draw[fermion] (4.828,0) node[right] {\small $X^-_1\left(\dfrac{1}{2}\right)$} -- (3.414,-1.414);
    \draw[fermion] (3.414,-1.414) -- (4.828,-2.828) node[right]
    {\small $X^0_3\left(-\dfrac{1}{2} \right)$};
  \end{tikzpicture}
  \hspace{1cm}
  \begin{tikzpicture}[scale=0.75]
    \draw[fermion] (2,0) -- (0,0) node[left] {\small $X^0_0\left(-\dfrac{1}{2}\right)$} ;
    \draw[fermion] (3.414,1.414) node[right] {\small $X^-_1\left(\dfrac{1}{2}\right)$} -- (2,0);
    \draw[scalar] (2,0) -- (3.414,-1.414);
    \draw[fermion] (3.414,-1.414) -- (4.828,0) node[right] {\small $X^+_1\left(\dfrac{1}{2}\right)$};
    \draw[fermion] (4.828,-2.828) node[right] {\small $X^0_3\left(\dfrac{1}{2}\right)$} -- (3.414,-1.414);
  \end{tikzpicture}
  \caption{Diagrams for the decay of a fermionic $X_0^0$.  Note the different spin structures.\label{fig:fdiags}}
\end{figure}
To see this in practice, we consider our toy model with the following
modification.  We let $X_0$, $X_1$ and $X_3$ now be fermions, while
keeping $Y$ a scalar.  We further suppose the following interaction
Lagrangian:
\begin{equation}
  \label{eq:Lint-ferm}
  - \mathcal{L} = \lambda_1 Y^+ \overline{X}_0^0 P_L X_1^- + \lambda_2
  Y^+ \overline{X}_3^0 P_L X_1^- + {\rm h.c.} \; ,
\end{equation}
where $P_L$ is the left-chirality projector.  The decay of
$X_0$ is mediated by the trivially different diagrams illustrated in
Fig.~\ref{fig:fdiags}.  The helicity structure of each decay is fixed
by the chiral structure as indicated by the listed helicities.  The
problem is now clear: there is no interference because the two
diagrams have different helicities of the $X_0$ and $X_3$.

For massive fermions, chirality is not the same as helicity and an
interference term can be generated but the CP asymmetry will suffer a
chiral suppression by the mass of the $X_3$ compared to the $X_0$.
Explicitly, the helicity-summed interference term in the squared
amplitude is given by
\begin{equation}
  \label{eq:int-chi}
  (|\mathcal{M}|^2)_{\rm int} = 2 {\rm Re}\left[\frac{-\lambda_1^{*2}
      \lambda_2^2\, m_0 m_3 \,p_+ \cdot p_-}{(q_{23}^2 - m^2 + i \Gamma m) (q_{13}^2 - m^2 - i \Gamma m)}\right] \; .
\end{equation}
Note that $m_0$ and $m_3$ have to be Majorana masses, as the direction of fermion flow has to be reversed together with the chirality.

Up to phase-space effects, the rate of $X_0$ decay in this mode is set
by the scale $m_0$.  To get a maximal CP-violating asymmetry, we would
like to then have $m_3 \sim m_0$.  On the other hand, the region of
parameter space where this relation is true, in addition to being
quite small and non-generic, is precisely the region
where the full rate of $X_0 \to X_1^+ X_1^- X_3$ faces a phase-space
suppression.  We conclude that CP-odd observables of the sort we are
studying will always face suppression when the couplings
involved in the decay are chiral.  There remain many spin
configurations that do not suffer such a suppression.  In addition to
cases where the decay vertices are vector- or pseudovector-like, any
situation where the neutral particles are bosons are not in danger of
chiral suppression.

\begin{figure}[tb]
  \centering
  \begin{tikzpicture}[scale=0.85]
    \draw[fermion] (0,0) node[left] {$\chi_2^0$} -- (2,0);
    \draw[fermion] (2,0) -- (3.414,1.414) node [above right] {$\mu^-$};
    \draw[scalar-ch] (3.414,-1.414) -- node[above right] {$\tilde{\mu}$}  (2,0);
    \draw[fermion] (4.828,0) node[right] {$\mu^+$}  -- (3.414,-1.414);
    \draw[fermion] (3.414,-1.414) -- (4.828,-2.828) node[below right] {$\chi_1^0$} ;
 \end{tikzpicture}\hspace{1.5cm}
   \begin{tikzpicture}[scale=0.85]
    \draw[fermion] (2,0) -- (0,0) node[left] {$\chi_2^0$};
    \draw[fermion] (3.414,1.414) node [above right] {$\mu^+$} --  (2,0);
    \draw[scalar-ch] (2,0) -- node[above right] {$\tilde{\mu}$} (3.414,-1.414);
    \draw[fermion] (3.414,-1.414) --  (4.828,0) node[right] {$\mu^-$} ;
    \draw[fermion] (4.828,-2.828) node[below right] {$\chi_1^0$} -- (3.414,-1.414);
  \end{tikzpicture}
  \caption{Feynman diagrams for the CP-violating decay of a heavy
    neutralino in a supersymmetric example model.  $\chi_i^0$ is the
    $i$th neutralino.  $\tilde{\mu}$ is a smuon.}
  \label{fig:susy-eg-dec}
\end{figure}
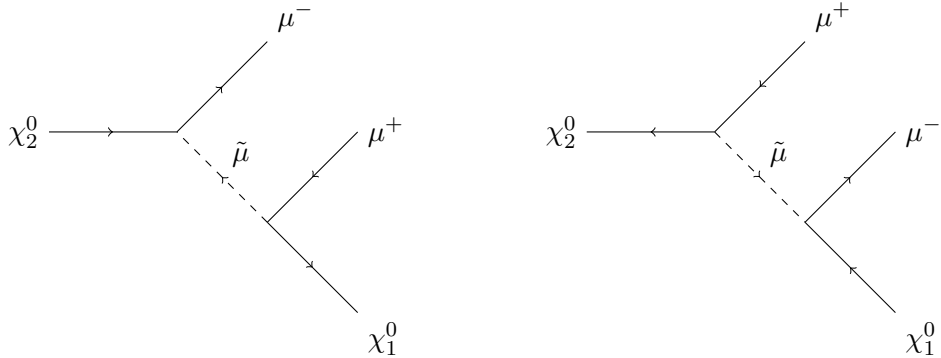

To see how much of an issue this is, we briefly discuss the
possibility of observing CP-violation in supersymmetric decays.  In
particular, we would like to consider pair production of heavy
neutralinos and the decay illustrated in Fig.~\ref{fig:susy-eg-dec}.
If the smuon is entirely left-handed, then the couplings are exactly
the $SU(2)$ and $U(1)$ gauge couplings, up to a CP-violating phase due
to the phases of the gaugino masses.  The only physical phase comes
from the combination $M_1 M_2^*$.

This particular decay, like most in weakly 
coupled theories, faces an additional challenge: the CP-violating
asymmetries are all suppressed by the small width of the intermediate
resonance, in this case the smuon.  This suppression combines with the
chiral suppression to give a very small ideal CP-violating asymmetry
\begin{equation}
  \label{eq:asym-susy}
  \mathcal{A}_{CP}^{\rm RF} = 0.010 \; ,
\end{equation}
where we assume a maximal CP-violating phase, no neutralino mixing and a
mass spectrum of $m_{\chi_2^0}=300$ GeV, $m_{\tilde{\mu}}= 250$ GeV,
and $m_{\chi_1^0}=200$ GeV. This is likely too small to get a
significant measurement at the LHC.

We would like to mention in passing that it is possible to overcome the 
chiral suppression even in case of chiral fermion couplings. One example 
are decays of the type $X^+_0 \to X_1^+ X_2^+ X_3^-$, where one would simply 
construct a standard charge asymmetry. Another example is discussed in more 
detail in the next section.

\section{Different Intermediate Resonances}
\label{sec:non-ident-interm}

In this section, we explore the possibility of expanding our
observable to the case where there are additional diagrams with new
resonances.  If the resonances have the same quantum numbers, then
this precisely describes CP-violation from mixing.  Depending on the
value of the mixing parameters $\Delta m$ and $\Delta \Gamma$, the
CP-even phase will arise predominantly from either the absorptive or
dispersive part of the mixing matrix.  We do not discuss this
possibility further.

Another possibility is that the two resonances have different quantum
numbers. As an example, we consider the Standard Model extended by 
a right-handed heavy Majorana neutrino $N_1$, having mass at the
weak scale and zero hypercharge. The Lagrangian is given by
\begin{equation}
\mathcal{L}_{N_1} = i \overline{N}_1 \dslash N_1 - 
\left( \frac{1}{2}\overline{N}_1 m_{N_1} N_1^C + {\rm {h.c.}} \right)
- \left( Y_\nu  \overline{N}_1 {\widetilde \phi}^\dagger l_L + {\rm {h.c.}} \right) \; , 
\end{equation}
where $m_{N_1}$ is the Majorana mass of $N_1$ and $Y_\nu$ is the neutrino Yukawa coupling.
%

\begin{figure}[tb]
  \centering
    \begin{tikzpicture}[scale=0.85]
    \draw (0,0) node[left] {$N_1$} -- (2,0);
    \draw[fermion] (2,0) -- (3.414,1.414) node [above right] {$\mu^-$};
    \draw[gauge] (3.414,-1.414) -- node[above right] {$W^+$}  (2,0);
    \draw[fermion] (4.828,0) node[right] {$\mu^+$}  -- (3.414,-1.414);
    \draw (3.414,-1.414) -- (4.828,-2.828) node[below right] {$\nu_m$} ;
 \end{tikzpicture}
 \hspace{1.5cm}
 \begin{tikzpicture}[scale=0.85]
    \draw (2,0) -- (0,0) node[left] {$N_1$};
    \draw (3.414,1.414) node [above right] {$\nu_m$} --  (2,0);
    \draw[gauge] (2,0) -- node[above right] {$Z$} (3.414,-1.414);
    \draw[fermion] (3.414,-1.414) --  (4.828,0) node[right] {$\mu^-$} ;
    \draw[fermion] (4.828,-2.828) node[below right] {$\mu^+$} -- (3.414,-1.414);
  \end{tikzpicture}
\caption{Feynman diagrams leading to CP-asymmetry in the decay of $N_1$ with 
different resonances.}
  \label{fig:diff-res-decay}
\end{figure}
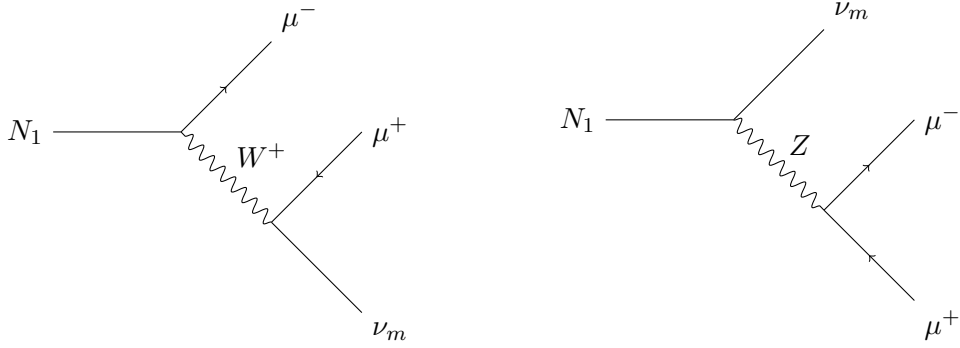

This model belongs to a generic class known as the Type-I seesaw
model.  This heavy neutrino will then mix with the active left-handed
light neutrinos $\nu_m$ ($m = 1,2,3$) and can decay via $Z$, $W^\pm$
or Higgs, according to the Lagrangian
\begin{equation}
- \mathcal{L}_{N_1,{\rm int}} = \frac{g}{\sqrt{2}} \lambda_1 \; \overline{N_1} \gamma^\mu P_L \ell W_\mu^+  
+ \frac{g}{2 c_w} \lambda_2  \; \bar\nu \gamma^\mu  P_L N_1 Z_\mu  
+ \frac{1}{\sqrt{2}} \lambda_3 \; \overline{\nu} P_R\, N_1 \,h + {\rm {h.c.}} \; ,
\end{equation}
where $\lambda_{1,2,3}$ are complex numbers in general that depend on the mixing
of $N_1$ with the light active neutrinos and $c_w$ is the cosine of the weak
mixing angle. $\lambda_3$ depends also on the neutrino Yukawa coupling $Y_\nu$.
Such an $N_1$ decay produces a pair of oppositely charged leptons $\ell^\pm$
and a light neutrino $\nu_m$, as can be read off from the diagrams
contributing to the heavy neutrino decay via $Z$ and $W^+$ that are
shown in Fig.~\ref{fig:diff-res-decay}.  Here we neglect the Higgs mediated diagram
as it receives a further suppression by two powers of Yukawa couplings.

In case of $N_1$ decays, we expect to have a CP-asymmetry from two sources: 
difference in the virtuality of the intermediate resonances due to 
different outgoing particles and also due to different masses and widths 
of the intermediate resonances. 
In most part of the parameter space, the CP-violation is small 
because it is suppressed either by the fact that the resonances cannot
be not sufficiently close to on-shell in both diagrams or by the
large mass splitting between the $W^+$ and the $Z$ and their 
small widths.  However, if the mass of $N_1$ is close to the
masses of both the $W$ and $Z$, there can be a significant
CP-asymmetry.  The intermediate particles must be off-shell to get a
non-vanishing CP-even phase, but it is advantageous to have them be
off-shell only by a small amount.  Thus, we choose $m_W < m_{N_1} <
m_Z$. The asymmetry also depends on $|\lambda_{1,2}|$ as well as the
relative phase between $\lambda_1$ and $\lambda_2$. Hence to have
maximum CP-asymmetry we choose ${\rm arg}(\lambda_1^\ast \lambda_2) =
\pm \pi/2$.  Finally, choosing an $N_1$ mass of 90 GeV we get
\begin{equation}
\label{eq:asym-N}
\mathcal{A}_{\rm CP}^{\rm RF} = 0.046 \; .
\end{equation} 
The asymmetry decreases sharply as we move away from the intermediate
particle mass range. In eq.~(\ref{eq:asym-N}) the couplings are chosen
as $|\lambda_1| = 0.04$, $|\lambda_2| = 0.3$. Note that since the
external particles are fermions we need to worry about chiral
suppression, but it is not very significant here since the two
intermediate particles couple significantly to the left handed
components.

Because of the relatively small widths of the intermediate $W$ and $Z$
particles, the momentum smearing effects at the detectors may have a 
large impact on the measured asymmetries in this scenario. Considering
the parametrization of transverse momentum smearing as given in
eq.~(\ref{eq:muon-res}), we find that smearing will be smaller than
the $W$ and $Z$ widths for $p_T \lsim 100$ GeV, and so we expect
negligible smearing effect for a $N_1$ of mass 90 GeV decaying at
rest.

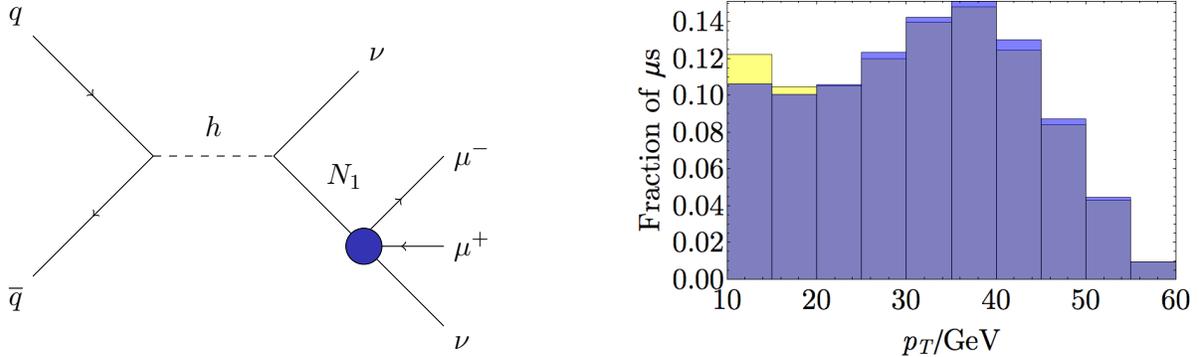
\begin{figure}[tb]
\begin{center}
 \begin{tikzpicture}[scale=0.8]
    \draw[fermion] (-2.0,2.0) node [above left] {$q$} -- (0,0) ;
    \draw[fermion] (0,0) -- (-2.0,-2.0)node [below left] {$\overline q$}  (2,0);
    \draw[scalar] (0,0) node at (1,0.5) {$h$} -- (2,0);
    \draw (2,0) -- (3.414,1.414) node [above right] {$\nu$};
    \draw (3.414,-1.414) -- node[above right] {$N_1$}  (2,0);
    \draw[fermion] (3.414,-1.414)  -- (4.828,0) node[right] {$\mu^-$};
    \draw[fermion] (4.828,-1.5) node[right] {$\mu^+$}  -- (3.414,-1.5);
    \draw (3.414,-1.414) -- (4.828,-2.828) node[below right] {$\nu$} ;
    \draw[fill=myblue] (3.5,-1.5) circle (0.3cm);
 \end{tikzpicture}
\hspace{1.5cm}
\includegraphics[width=0.45\textwidth]{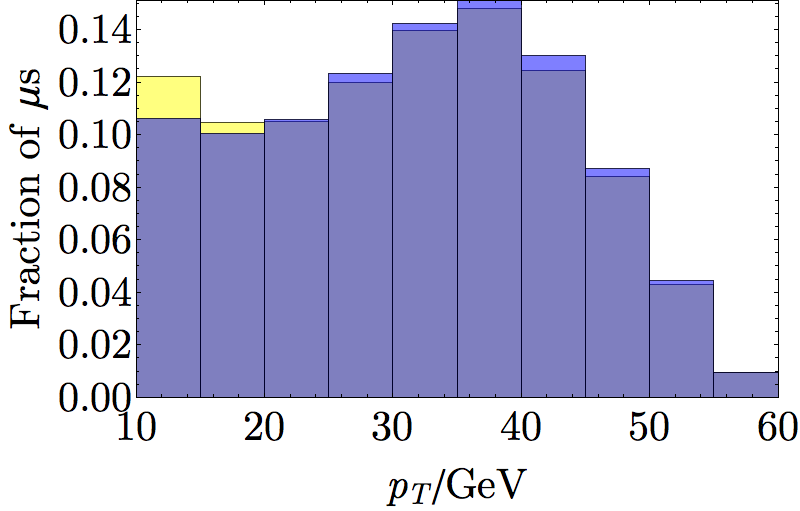}
\caption{Left: Feynman diagram leading to CP-asymmetry in 
the decay of $N_1$ produced from the resonantly produced Higgs at LHC. 
Right: $p_T$ distributions of $\mu^-$ (yellow) and $\mu^+$ (blue)
produced in this process with the overlap in purple.}
  \label{fig:LHC-h-N1}
\end{center}
\end{figure}
The situation becomes much more complicated at a collider when the
parent heavy-neutrino is produced in a high energy collision.
Here we consider an example where $N_1$ is produced via Higgs decay as
shown in the left figure of Fig.~\ref{fig:LHC-h-N1}. Since
the Higgs is dominantly produced as a resonance at LHC, we consider
the resonant Higgs production at the leading 
order for illustrative purpose.  The subsequent decay of the
Higgs produces $N_1$.  We consider a Higgs of mass 120
GeV and choose $\lambda_3$ to be real.  The transverse momentum
asymmetry, for the above chosen parameters, becomes 
\beq
\mathcal{A}_{\rm CP}^{p_T} = 0.028 \; ,
\eeq
which shows a suppression by a factor of $\sim 0.6$ compared to the
ideal asymmetry in eq.~(\ref{eq:asym-N}). The suppression is solely
due to the fact that the parent $N_1$ is not at rest in the transverse
direction. In this example we do not consider the $N_1$ production
channels via $W$ or $Z$, since there will then be a CP-odd phase in
the production as well, and the final asymmetry will be some
combination of the production phase and the decay phase.

Since Higgs is produced resonantly, it has negligible 
transverse momentum and $\mu^\pm$ produced finally 
in the $N_1$ decay has $p_T \lsim 60~{\rm GeV}$, as can be seen from 
the right plot of Fig.~\ref{fig:LHC-h-N1}.  Smearing effects are thus
expected to be negligible, as we have verified using Monte Carlo
generated events.

\section{Conclusions}
\label{sec:conclusions}

In this paper, we have studied in detail CP violating momentum
asymmetries arising in three body decays with the goal of determining
their usefulness at the LHC. We concentrated on cases where one of the
final state particles cannot be detected. Our main result is that
lab-frame transverse momentum asymmetries can be constructed that can
be used to probe CP violation.

We have analyzed three different production mechanisms, namely
resonant production, pair production and production in decay.  In each
case we have identified the relevant transverse momentum asymmetries
and studied their robustness in a realistic hadron collider
environment.  The present analysis therefore greatly increases the
range of models in which CP violating observables of the sort
discussed in~\cite{Berger:2011wh} can be used. Our results indicate
that there are significant constraints on the domain where these
observables are practically useful. There are however regions of
parameters space where they can be large. This is generally the case
when the resonance is wide and the production mechanism does not
introduce a very large transverse boost. More generally, we think of
momentum asymmetries as complementary to triple products: the better
choice depends on the model and its parameters. Clearly, we 
should be ready for any new physics model that nature may impose on us.

To conclude, new sources of CP violation are generally expected to 
emerge with new physics at the TeV scale. Constructing CP-odd observables 
that allow to experimentally disentangle them at the LHC is a challenging task. 
The momentum asymmetries studied in the present paper provide an alternative 
route to measure CP violating phases in cases where well-known observables, 
such as triple product asymmetries, cannot be constructed. We therefore expect 
CP violating momentum asymmetries to be relevant in a wide variety of new 
physics scenarios.

\section*{Acknowledgements}

We are grateful for helpful discussions with Jay Wacker and Natalia
Toro during the completion of this work. This
work was supported by the U.S. National Science Foundation
through grant PHY-0757868 and CAREER grant No.\ PHY-0844667.

\end{document}